\documentclass[pra,showpacs,preprint,preprintnumbers,amsmath,amssymb]{revtex4}

\usepackage{graphicx}

\begin{document}

\title{Experimental entanglement distillation using Schmidt projection on 
multiple qubits}

\author{Taehyun Kim\footnote{Electronic address: thkim@alum.mit.edu} and Franco 
N. C. Wong}
\affiliation{Research Laboratory of Electronics, Massachusetts Institute of 
Technology, Cambridge, Massachusetts 02139, USA}

\begin{abstract}
Entanglement distillation is a procedure for extracting from one or more pairs 
of entangled qubits a smaller number of pairs with a higher degree of 
entanglement that is essential for many applications in quantum information 
science.  Schmidt projection is a unique distillation method that can be applied 
to more than two pairs of entangled qubits at the same time with high 
efficiency.  We implement the Schmidt projection protocol by 
applying single-photon two-qubit quantum logic to a partially hyperentangled 
four-qubit input state. The basic elements of the procedure are characterized 
and we confirm that the output of Schmidt projection is always maximally 
entangled independent of the input entanglement quality and that the measured 
distillation efficiency agrees with theoretical predictions.  We show that our 
implementation strategy can be applied to other quantum systems such as trapped 
atoms and trapped ions.
\end{abstract}

\maketitle
\section{INTRODUCTION}
Shared entanglement is an essential resource in many quantum information 
processing applications such as quantum teleportation \cite{Teleportation}, 
quantum computation \cite{KLM,QTeleportationComputing,OneWayQComputer}, quantum 
cryptography \cite{QKDEntanglement}, and quantum repeaters \cite{QRepeater}.
These applications generally require that two remote parties maintain a very 
high degree of entanglement in order for their protocols to work properly.  
Therefore, the generation of maximally entangled states such as the Bell states 
is an essential task in quantum information science.   However, even if one can 
initially generate maximally entangled states with ease, they often become 
non-maximally entangled or partially mixed at the remote locations because of 
dissipation and decoherence through interactions with the environment.  While 
each application has its own level of tolerance to entanglement degradation, 
most applications would be practically unusable without some kind of procedure 
to restore the degraded entanglement to its initial maximally entangled 
state.  

The need for most quantum information processing applications to overcome 
unavoidable loss of entanglement has led to several methods for entanglement 
restoration \cite{Concentration, PurificationCNOT, PurificationParity1, 
LocalFiltering, NoPostSelection, Procrustean, PurificationParity2, Parity1, 
Parity2}.   These techniques typically seek to improve the quality of 
entanglement between a pair of entangled qubits by selecting, filtering, and 
applying unique processing algorithms to a larger number of lower quality 
entangled qubits.  Entanglement restoration provides a path to 
utilizing entangled qubits that would otherwise be dicarded because their
entanglement quality is below the threshold necessary for 
specific applications.  The cost of entanglement restoration is that many pairs of the lower 
quality entangled qubits must be generated and processed, which requires more 
resources and longer times for each application processing step.  It is 
therefore of interest to choose a method that is efficient in terms of resources 
and processing required to achieve a certain level of entanglement quality.  
Among these methods, Schmidt projection (SP) \cite{Concentration} has the  
interesting property that it can be applied to more than two pairs of entangled 
qubits at the same time and that the distillation efficiency can approach unity 
when applied to a large number of initial pairs.  Despite these advantages, 
Schmidt projection is considered difficult to implement because it requires 
simultaneous collective measurements on multiple qubits.
 
In this work, we illustrate the working principle of Schmidt projection 
\cite{Concentration} by applying the technique to two pairs of partially 
entangled qubits.  First, we use the concept of physical mapping to encode the 
two pairs of entangled qubits into four physically distinguishable states of two 
hyperentangled photons.  The two photons are entangled in both polarization and 
momentum degrees of freedom so that the four qubits can be distinguished by 
their photon, polarization, and momentum.  By hosting two qubits per photon 
carrier, we eliminate the problem of bringing the two qubits together to overlap 
completely in time for quantum logic gate operations.  Moreover, the two types 
of qubits can be manipulated deterministically using single-photon two-qubit 
(SPTQ) quantum logic \cite{PCNOT,SPTQSWAP}.  The SP protocol is carried out 
efficiently using SPTQ quantum gates to produce a single maximally-entangled 
pair of qubits.  In Sect.~II we describe the basic idea of the SP protocol and 
show how this protocol can be implemented with hyperentangled photon pairs.  
Section~III describes our experimental implementation including the generation 
of  partially hyperentangled photons and the use of SPTQ logic to perform the 
Schmidt projection.  The resulting maximally polarization-entangled photons are 
characterized and the effectiveness and efficiency of the protocol are analyzed 
and discussed in Sect.~IV, and finally we summarize and show how the concept of 
physical mapping can be applied to trapped-atom and trapped-ion systems.

\section{THEORETICAL BACKGROUND}
Before we describe how the Schmidt projection protocol can be implemented in our 
experiment, it is useful to clarify the sometimes confusing terminology in 
entanglement restoration.  Given one or more pairs of entangled qubits one can 
extract from them a smaller number of pairs with a higher degree of entanglement 
(or the same number of pairs with less than unity probability) by using local 
operations and classical communication.  This process of improving the amount of 
entanglement is called entanglement distillation.  In the literature, this 
process is sometimes called entanglement purification or entanglement 
concentration.  In this work, we follow the convention given by 
Ref.~\cite{LocalFiltering}, in which entanglement purification refers to the 
process of enhancing the purity of a mixed state by extracting a number of purer 
entangled states from a larger number of less pure entangled pairs.  
Entanglement concentration is the process for increasing both the degree of 
entanglement and the purity of a given initial state.

To be more specific, we use the definition suggested by Bennett {\em et al.} 
\cite{Concentration} to quantify entanglement.  For a partially entangled pure 
state $|\psi\rangle$ shared by Alice ($A$) and Bob ($B$), the von Neumann 
entropy of the partial density matrix gives a measure of the entanglement 
$E(\psi)$:
\begin{equation}
E(\psi)=-{\rm Tr}(\rho_A {\rm log}_2 \rho_A)=-{\rm Tr}(\rho_B {\rm log}_2 
\rho_B)\,,
\label{Eq:EntanglementOfPureState}
\end{equation}
where $\rho_A (\rho_B)$ is the partial trace of $|\psi\rangle\langle\psi|$ over 
subsystem $B$ ($A$). In general, the entanglement measure is bounded by $0 \leq 
E(\psi) \leq \log_2 d$ for a bipartite qudit system in a $2d$-dimensional 
Hilbert space.  A natural extension of the definition to a mixed state with a 
density matrix $\rho$ is the entanglement of formation $E(\rho)$, which is the 
smallest expectation value of entanglement $E$ of any ensemble of pure states 
realizing $\rho$ \cite{MixedStateEntanglement}.  In other words, to calculate 
the entanglement of formation, one needs to consider all possible decompositions 
of $\rho$, that is, all ensembles of states $|\psi_i\rangle$ with probabilities 
$p_i$ satisfying $\rho=\Sigma_i p_i |\psi_i\rangle\langle\psi_i|$.  Then the 
entanglement of formation is defined as $E(\rho)={\rm min} \Sigma_i p_i 
E(\psi_i)$, which in general is difficult to calculate. However, for a bipartite 
qubit system, an explicit formula for entanglement of formation exists 
\cite{EOF}.

Most of the protocols for enhancing $E$ can only be applied to one 
\cite{LocalFiltering,Procrustean} or two pairs of entangled qubits at a time 
\cite{Parity1,Parity2,NoPostSelection} and therefore have a 
fixed distillation efficiency independent of the number of initial pairs.  In 
contrast, Bennett {\it et al.} showed that Schmidt projection can be applied to 
$n$ pairs of partially entangled pure state with $E(\psi)$ to extract $n\cdot 
E(\psi)$ pairs of maximally entangled Bell state in the limit of large $n$ 
\cite{Concentration}, and this property is used to justify the definition of $E$ 
in Eq.~(\ref{Eq:EntanglementOfPureState}) as a measure of the degree of 
entanglement.  Entanglement $E(\psi)$, and hence the SP technique, forms the 
basis of or is related to other common measures such as entanglement of 
formation, concurrence \cite{Concurrence,ConcurrenceMeasure}, and tangle 
\cite{Tangle}.  Despite its role in the theoretical foundation of quantum 
information science, Schmidt projection has not been experimentally demonstrated 
due to the difficulty of performing the required collective measurements.

To understand how Schmidt projection works, consider two pairs of identically 
entangled state, $|\psi\rangle_{AB}=\cos\theta|0\rangle_A\otimes|0\rangle_B + 
\sin\theta|1\rangle_A\otimes|1\rangle_B$, which can be described by
\begin{eqnarray}
\lefteqn{|\psi\rangle_{A_1 B_1}\otimes|\psi\rangle_{A_2 B_2}} \nonumber \\
& &=\cos^2\theta|0\rangle_{A_1}|0\rangle_{A_2}|0\rangle_{B_1}|0\rangle_{B_2}
+\cos\theta\sin\theta 
|0\rangle_{A_1}|1\rangle_{A_2}|0\rangle_{B_1}|1\rangle_{B_2} \nonumber \\ 
& &+\sin\theta\cos\theta 
|1\rangle_{A_1}|0\rangle_{A_2}|1\rangle_{B_1}|0\rangle_{B_2}
+\sin^2\theta |1\rangle_{A_1}|1\rangle_{A_2}|1\rangle_{B_1}|1\rangle_{B_2}\,, 
\label{Eq:Expansion} \\
& &\equiv \cos^2\theta|0\rangle_A|0\rangle_B +\sin^2\theta 
|3\rangle_A|3\rangle_B 
+\cos\theta\sin\theta (|1\rangle_A|1\rangle_B+|2\rangle_A|2\rangle_B)\,.
\label{Eq:NumberNotation}
\end{eqnarray}
In Eq.~(\ref{Eq:NumberNotation}) we replace the binary representation 
of Alice's qubits and Bob's qubits by qudit notation (decimal number). For two 
entangled pairs, the task of collective measurements is reduced to projecting 
terms with the same coefficient ($\cos\theta\sin\theta$) whose output form is 
that of a maximally entangled state.   Equation~(\ref{Eq:NumberNotation}) shows 
that each of the three terms is locally orthogonal to each other, and therefore 
the result of Alice's local projection is perfectly correlated to that of Bob's 
local projection, thus making post-processing and the associated classical 
communication unnecessary.  In contrast, the non-Schmidt projection methods 
used in Ref.~\cite{Parity1,Parity2} require that Alice and Bob compare their 
measurement results over a classical channel and change the phase based on their 
measurements.  

Schmidt projection is most useful when it is applied to $n>2$ pairs of 
$|\psi\rangle_{AB}$ that has the initial tensor product state given by
\begin{eqnarray}
\lefteqn{|\psi\rangle_{A B}^{\otimes n}} \nonumber \\
& & =\left(\cos\theta|0\rangle_{A_1}|0\rangle_{B_1} + 
\sin\theta|1\rangle_{A_1}|1\rangle_{B_1}\right)\otimes \cdots \otimes 
\left(\cos\theta|0\rangle_{A_n}|0\rangle_{B_n} + 
\sin\theta|1\rangle_{A_n}|1\rangle_{B_n}\right) \nonumber \\
& & =\cos^n\theta|0\rangle_{A_1} \ldots |0\rangle_{A_n}|0\rangle_{B_1} \ldots 
|0\rangle_{B_n} \nonumber \\
& & +\cos^{n-1}\theta\sin\theta ( |0\rangle_{A_1}\ldots|0\rangle_{A_{n-
1}}|1\rangle_{A_n}
|0\rangle_{B_1}\ldots|0\rangle_{B_{n-1}}|1\rangle_{B_n}+\cdots \nonumber \\
& & 
\;\;\;\;\;\;\;\;\;\;\;\;\;\;\;\;\;\;\;\;\;\;\;\;+|1\rangle_{A_1}|0\rangle_{A_2}
\ldots|0\rangle_{A_n}
|1\rangle_{B_1}|0\rangle_{B_2}\ldots|0\rangle_{B_n} ) \nonumber \\ 
& & + \cdots +\sin^n\theta|1\rangle_{A_1} \ldots |1\rangle_{A_n}|1\rangle_{B_1} 
\ldots |1\rangle_{B_n} .
\label{Eq:PowerN}
\end{eqnarray}
Among the $2^n$ terms in Eq.~(\ref{Eq:PowerN}), there are $n+1$ distinct Schmidt 
coefficients ($\cos^n\theta,$ $\cos^{n-1}\theta\sin\theta, ...,\sin^n\theta$), 
and each group with the same coefficient $\cos^{n-k}\theta\sin^k\theta$ has 
$(^n_k)$ terms that are maximally entangled, for $n > k > 0$ 
\cite{Concentration}.  Alice and Bob project this initial state into one 
of $n-1$ orthogonal subspaces to extract the associated maximally-entangled 
states.  The extracted state can be directly used for faithful teleportation in 
a $(^n_k)$-dimensional or smaller Hilbert space, or, alternatively, one can 
efficiently convert this state into a tensor product of Bell states, as 
suggested by Bennett {\it et al.} \cite{Concentration}.

In our experiment, we demonstrate a proof-of-principle implementation of Schmidt 
projection by applying the SP protocol to two pairs of entangled qubits such as 
those given by Eq.~(\ref{Eq:NumberNotation}).  We choose to implement the 
four-qubit SP protocol with a pair of partially hyperentangled photons as the 
initial state given by
\begin{eqnarray}
|\psi_{P}\rangle\otimes|\psi_{M}\rangle 
&= \left( \cos \theta_P |V\rangle_A 
|V\rangle_B + \sin\theta_P |H\rangle_A |H\rangle_B \right) \nonumber \\
&\otimes\left(\cos\theta_M |L\rangle_A |L\rangle_B + \sin\theta_M |R\rangle_A 
|R\rangle_B \right)\,, \label{Eq:Hyperentangled1}
\end{eqnarray}
where subscript $P$ ($M$) refers to the polarization (momentum) degree of 
freedom, $H$ ($V$) represents horizontal (vertical) polarization, and $L$ ($R$) 
is the left (right) path shown in Fig.~\ref{Fig:DistillGeneration}.  Schmidt 
projection requires that the two input qubit pairs are identically entangled, 
$\theta_P=\theta_M=\theta$, which we are able to set by the method described in 
the next section.  With this setting, Eq.~(\ref{Eq:Hyperentangled1}) can be 
expanded as
\begin{eqnarray}
& & \cos^2\theta |VL\rangle_A |VL\rangle_B+\sin^2\theta |HR\rangle_A 
|HR\rangle_B \nonumber\\
& & + \cos\theta \sin\theta\left( |VR\rangle_A |VR\rangle_B + |HL\rangle_A 
|HL\rangle_B \right).\label{Eq:Hyperentangled2}
\end{eqnarray}
In Eq.~(\ref{Eq:Hyperentangled2}) the SP protocol is applied by projecting the 
term associated with the coefficient $\cos\theta \sin\theta$ that can be 
realized by transmitting the $V$-polarized photon in the $R$ path and the 
$H$-polarized photon in the $L$ path by both Alice and Bob.  

\section{EXPERIMENTAL SETUP}
The use of hyperentangled photons for demonstrating a four-qubit SP protocol has 
a number of experimental advantages.  Only a pair of hyperentangled photons is 
needed to produce two pairs of entangled qubits, one pair entangled in the 
polarization mode and the other pair in the momentum (or path) mode.  Therefore, 
the photon serves as a carrier for the polarization and momentum qubits that can 
be manipulated with single-qubit and two-qubit quantum gates using deterministic 
single-photon two-qubit quantum logic gates \cite{PCNOT,SPTQSWAP}.  We have 
previously demonstrated deterministic SPTQ quantum logic gates such as 
polarization-{\sc cnot} \cite{PCNOT}, momentum-{\sc cnot} and {\sc swap} 
\cite{SPTQSWAP} using linear optical components.  The SPTQ implementation is 
simple and offers a useful platform for few-qubit quantum information processing 
tasks.  In this section, we describe the method for generating partially 
hyperentangled photon pairs with adjustable degrees of entanglement using a 
bidirectionally pumped polarization Sagnac interferometer (PSI).  We then detail 
the experimental implementation of the SP protocol for extracting a pair of 
maximally-entangled qubits from the initial partially hyperentangled four-qubit 
state using SPTQ quantum logic.  The setup for verifying the input and output 
states of the distillation procedure will be described.

\subsection{Generation of partially hyperentangled photon pairs}
\begin{figure}
  \begin{center}
    \includegraphics[width=12cm]{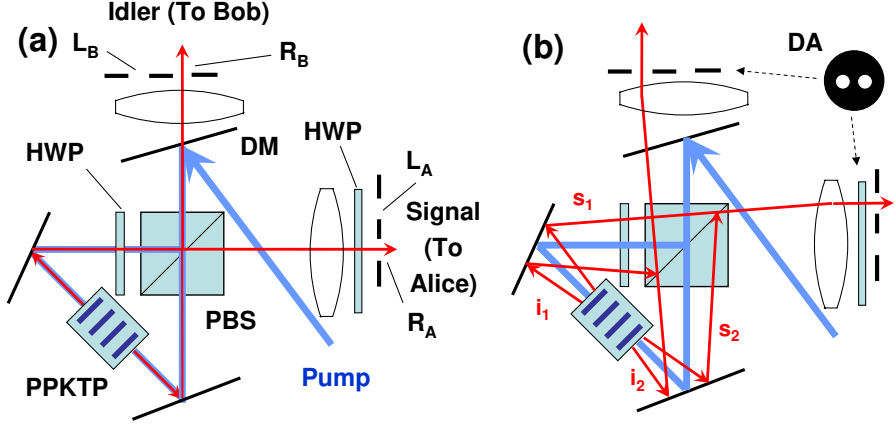}  
  \end{center}
  \caption{(Color online) Hyperentangled photon pair source in a PSI 
configuration, showing generation of polarization-entangled photon pairs in (a) 
collinear output modes along $R_A$ and $R_B$ and (b) non-collinear output modes 
along $L_A$ and $L_B$. The PSI consists of the crystal, two flat mirrors, a 
dual-wavelength half-wave plate (HWP), and a dual-wavelength polarizing beam 
splitter (PBS).  The hyperentangled output state is a superposition of these two 
output modes each with a different probability amplitude. The two HWPs are used 
to rotate the polarizations by 90$^\circ$.  The subscripts 1 and 2 for the 
signal and idler outputs in (b) refer to the clockwise and counter-clockwise 
directions of the SPDC outputs, respectively. DM: dichroic mirror, DA: double 
aperture mask.}
  \label{Fig:DistillGeneration}
\end{figure}

To generate the hyperentangled state that are entangled in both polarization and 
momentum degrees of freedom, we utilized a polarization-entangled photon pair 
source based on spontaneous parametric down-conversion (SPDC)
in a type-II phase-matched periodically poled KTiOPO$_4$ (PPKTP) crystal in a 
bidirectionally pumped PSI configuration \cite{SagnacSource,SagnacSource2}.  The 
SPDC source was driven with $\sim$5 mW of a continuous-wave (cw) diode laser 
operating at a wavelength of 404.775 nm that was delivered to the PSI in a 
single-mode optical fiber. The 1-cm long PPKTP crystal was set at a temperature 
of $\sim$19\,$^\circ$C for nearly frequency-degenerate operation.  The signal 
and idler outputs at $809.55 \pm 0.5$ nm were detected through 1-nm interference
filters (IF) to restrict the SPDC output bandwidth.  More details of the PSI 
source can be found in Ref.~\cite{SagnacSource,SagnacSource2}.

Normally the PSI entanglement source is operated in a collinearly propagating 
geometry for high collection efficiency of the polarization-entangled photons.  
However, it can also be operated in a slightly non-collinear configuration to 
exploit the momentum correlation of SPDC outputs.  By collecting the 
non-collinear output modes from the PPKTP crystal the output photons are also 
momentum entangled \cite{MtmEntanglement,PCNOT}, as illustrated schematically in 
Fig.~\ref{Fig:DistillGeneration}. After the PSI outputs are collimated the 
momentum of the output mode can be identified equivalently by its path location 
that is defined by the double-aperture (DA) mask.  Unlike other 
hyperentanglement sources \cite{Hyperentangled1,Hyperentangled2,AVNExp2}, our 
experiment requires partial hyperentanglement with adjustable but equal amounts 
of entanglement in both polarization and momentum degrees of freedom, as 
indicated in Eq.~(\ref{Eq:Hyperentangled1}).  To control the degree of momentum 
entanglement ($\theta_M$), we aligned the DA mask at an asymmetric location
with respect to the pump beam axis so that the right aperture ($R_A$ or $R_B$) 
coincided with the pump axis as shown in Fig.~\ref{Fig:DistillGeneration}.  The  
diameter of each aperture of the DA mask was 1 mm, and the distance between the 
centers of two apertures was 2 mm.

Consider the collinear $R_A$ and $R_B$ output modes of 
Fig.~\ref{Fig:DistillGeneration}(a), which is the standard PSI configuration for 
generating polarization-entangled photons.  We showed in previous PSI 
experiments \cite{SagnacSource,SagnacSource2} that the output state is 
proportional to
\begin{equation}
|\psi\rangle_{R_A R_B} \equiv (\cos\theta_P 
\hat{a}_{H, R_A}^\dagger \hat{b}_{V, R_B}^\dagger + e^{i\phi} \sin\theta_P 
\hat{a}_{V, R_A}^\dagger \hat{b}_{H, R_B}^\dagger)|0\rangle\,,
\label{Eq:RARBTerm}
\end{equation}
where $\hat{a}_{V,R_A}^\dagger (\hat{b}_{H,R_B}^\dagger)$ is a 
$V$ ($H$) polarized photon creation operator for Alice's $R_A$ (Bob's $R_B$) 
path, and similarly for $\hat{a}_{V,R_B}^\dagger$ and $\hat{b}_{H,R_A}^\dagger$.  
The relative amplitude ($\theta_P$) and phase ($\phi$) of the output state are 
determined by the adjustable relative amplitude and phase between the $H$ and 
$V$ polarization components of the pump.  Similarly, for the non-collinear 
$L_A$--$L_B$ output modes of Fig.~\ref{Fig:DistillGeneration}(b), the output 
state is proportional to 
\begin{equation}
|\psi\rangle_{L_A L_B} \equiv (\cos\theta_P \hat{a}_{H, L_A}^\dagger \hat{b}_{V, 
L_B}^\dagger + 
e^{i\phi} \sin\theta_P \hat{a}_{V, L_A}^\dagger \hat{b}_{H, L_B}^\dagger) 
|0\rangle\,,
\label{Eq:LALBTerm}
\end{equation}
where $\theta_P$ and $\phi$ are the same as those in Eq.~(\ref{Eq:RARBTerm})
because they are generated by the same pump laser.  

For hyperentanglement, we consider the coherently driven outputs of all four 
paths and obtain a final state that is a superposition of the $|\psi\rangle_{R_A 
R_B}$ and $|\psi\rangle_{L_A L_B}$ outputs of Eq.~(\ref{Eq:RARBTerm}) and 
Eq.~(\ref{Eq:LALBTerm}), respectively.  In general, the collinear output modes 
($\hat{a}_{R_A}^\dagger,\hat{b}_{R_B}^\dagger$) and the non-collinear output 
modes ($\hat{a}_{L_A}^\dagger,\hat{b}_{L_B}^\dagger$) are excited with different 
probability amplitudes depending on the phase-matched emission angle 
\cite{ConeEntangle}.  Therefore, the four-path output is a partially 
hyperentangled state given by
\begin{eqnarray}
\lefteqn{\cos\theta_M|\psi\rangle_{L_A L_B} +\sin\theta_M|\psi\rangle_{R_A 
R_B}}\nonumber \\
& =& \cos\theta_M(\cos\theta_P|H L\rangle_A |V L\rangle_B + 
e^{i\phi}\sin\theta_P|V L\rangle_A |H L\rangle_B) \nonumber \\
& +& \sin\theta_M(\cos\theta_P|H R\rangle_A |V R\rangle_B + 
e^{i\phi}\sin\theta_P|V R\rangle_A |H R\rangle_B)\nonumber \\
& =& (\cos\theta_P|H_A\rangle |V_B\rangle + e^{i\phi}\sin\theta_P|V_A\rangle 
|H_B\rangle)\nonumber \\
& &\otimes(\cos\theta_M|L_A\rangle |L_B\rangle + \sin\theta_M|R_A\rangle 
|R_B\rangle)\,. \nonumber
\end{eqnarray}
By varying the crystal phase-matching temperature to change the solid angle of 
the SPDC emission cone, one can adjust the flux ratio between the collinear 
output mode and the non-collinear output mode, thereby controlling $\theta_M$.   
We set the pump relative phase such that the output phase $\phi = 0$, and the 
half-wave plate placed in Alice's path  in Fig.~\ref{Fig:DistillGeneration} was 
used to rotate Aice's output polarization by 90$^\circ$ to yield a partially 
hyperentangled output state given by Eq.~(\ref{Eq:Hyperentangled1}).
As an input state for Schmidt projection, we adjusted the pump relative 
amplitudes to set $\theta_P=\theta_M=\theta$.  

\subsection{Implementation of Schmidt projection}
\begin{figure}[ht]
  \begin{center}
    \includegraphics[width=12cm]{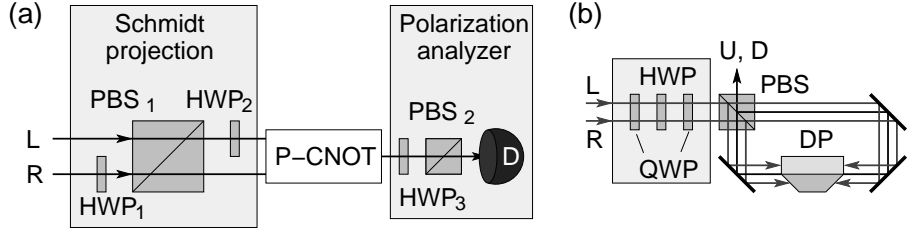}
  \end{center}
  \caption{(Color online) Entanglement distillation scheme for hyperentangled 
photons. Alice and Bob have the same setup. (a) Schmidt projection transmits 
only the two relevant terms, $|VR\rangle$ and $|HL\rangle$, of the initial state 
given in Eq.~(\ref{Eq:Hyperentangled2}).  {\sc p-cnot} gate combines two paths 
($R$, $L$) into a common path for final state extraction and for polarization 
state analysis. (b) {\sc p-cnot} gate with phase compensation. PBS: polarization beam splitter;
HWP: half-wave plate; QWP: quarter-wave plate; DP: dove prism.}
  \label{Fig:DistillSPSetup}
\end{figure}

We applied single-qubit and two-qubit gates of single-photon two-qubit quantum 
logic to implement Schmidt projection on the partially hyperentangled state of 
Eq.~(\ref{Eq:Hyperentangled1}), or its expanded form of 
Eq.~(\ref{Eq:Hyperentangled2}). The SP procedure is to extract from the initial 
state the maximally entangled term $|\psi\rangle_{SP} = |V R\rangle_A |V 
R\rangle_B + |H L\rangle_A |H L\rangle_B$, which is orthogonal to the other two 
terms in Eq.~(\ref{Eq:Hyperentangled2}).  Therefore, $|\psi\rangle_{SP}$ can be 
simply isolated by using a polarization beam splitter to separate the 
polarizations and choosing the appropriate beam paths.  However, the extracted 
state $|\psi\rangle_{SP}$ still contains four modes even though $V$ ($H$) is 
always associated with $R$ ($L$).  It is more useful to convert the extracted 
state to a more familiar form such as the polarization triplet state 
\begin{equation}
|\phi^+ \rangle = |H_A H_B + V_A V_B \rangle/\sqrt{2}\,.
\label{Eq:triplet}
\end{equation}
The transformation from $|\psi\rangle_{SP}$ to $|\phi^+ \rangle$ can be 
accomplished with a polarization controlled {\sc not} ({\sc p-cnot}) gate 
\cite{PCNOT} that converts $\psi\rangle_{SP}$ to $|\phi^+ \rangle \otimes 
|L\rangle$, thus bringing the momentum qubit to a fixed value, which we can 
ignore in the final output state $|\phi^+ \rangle$.  The polarization-entangled 
triplet final state is much easier to analyze using polarization correlation 
measurements than dealing with a state containing both $L$ and $R$ paths.  

Fig.~\ref{Fig:DistillSPSetup}(a) shows the apparatus for implementing the 
Schmidt projection method.  The key step is to project the initial state given 
by Eq.~(\ref{Eq:Hyperentangled2}) into a subspace composed of $|VR\rangle$ and 
$|HL\rangle$ for both Alice and Bob without destroying the qubits.  In 
principle, this can be accomplished by using one PBS in the $L$ path to transmit 
the $H$-polarized photon in the $L$ path and another PBS in the $R$ path but is 
rotated by 90$^\circ$ about the propagating axis to transmit the $V$-polarized 
photon in that path.  The first two terms in Eq.~(\ref{Eq:Hyperentangled2}) are 
then reflected.  In the actual setup of Fig.~\ref{Fig:DistillSPSetup} we modify 
this arrangement by using the HWP$_1$ to flip the polarization in the $R$ path 
before PBS$_1$ that transmits the desired subspace components in both paths.  
Instead of recovering the polarization state in the $R$ path after PBS$_1$, we 
flip the polarization state in the $L$ path with HWP$_2$, so that the output 
after Schmidt projection is $\cos\theta \sin\theta (|HR\rangle_A |HR\rangle_B + 
|VL\rangle_A |VL\rangle_B)$.  Note that HWP$_1$ and HWP$_2$ are essentially 
momentum-controlled {\sc not} ({\sc m-cnot}) gates described in 
Ref.~\cite{SPTQSWAP}.  We should point out that the arrangement of having one 
HWP in each path eliminates the need for path-length compensation between the 
two paths that is necessary if only one {\sc m-cnot} gate (i.e., a HWP) is used.  

To eliminate the path dependence of the Schmidt-projected state $|HR\rangle_A 
|HR\rangle_B + |VL\rangle_A |VL\rangle_B$, we fold the $L$-path output and the 
$R$-path output onto a common path by use of a {\sc p-cnot} gate, as indicated 
in Fig.~\ref{Fig:DistillSPSetup}(a).  The final output state becomes $|\phi^+ 
\rangle$ and we simplify the notation by removing the path designation.  The 
implementation of this {\sc p-cnot} gate is schematically shown in 
Fig.~\ref{Fig:DistillSPSetup}(b).  The {\sc p-cnot} gate consisted of a dove 
prism embedded in a polarization Sagnac interferometer that would rotate the 
incoming image ($L$ and $R$ beams) by $\pm90^\circ$ depending on the beam 
polarization \cite{PCNOT}.  The {\sc p-cnot} gate has the following mapping: 
$|HL\rangle\rightarrow|HU\rangle, |HR\rangle\rightarrow|HD\rangle, 
|VL\rangle\rightarrow|VD\rangle, |VR\rangle\rightarrow|VU\rangle$, where $U$ 
($D$) refers to the upside (downside) path.  For the purpose of relating to the 
usual $|0\rangle$ and $|1\rangle$ states, we identify $H$, $L$, and $U$ with 
$|0\rangle$, and $V$, $R$, and $D$ with $|1\rangle$.  Our {\sc p-cnot} 
implementation introduced a fixed amount of phase shifts depending on the input 
polarization \cite{SimAttackBB84} and they were compensated by adding a HWP 
sandwiched by two quarter-wave plates (QWPs) whose optic axes were tilted by 
45$^\circ$, as shown in Fig.~\ref{Fig:DistillSPSetup}(b).  This phase 
compensator was also utilized to add the necessary $\pi/2$ phase shift in our 
quantum state tomography measurements in the next Section.

To verify that the final output is the maximally-entangled polarization triplet 
$|\phi^+\rangle$, we analyzed the polarization correlation between Alice's and 
Bob's coincidence counts.  The polarization measurements were carried out with 
polarization analyzers each consisting of HWP$_3$, PBS$_2$, and a Si 
single-photon counter, as shown in Fig.~\ref{Fig:DistillSPSetup}(a). The 
measurement basis for the polarization analyzer was set by HWP$_3$.

\section{EXPERIMENTAL RESULTS}
Our implementation of the Schmidt projection protocol is best characterized by 
first demonstrating the required input hyperentangled state of 
Eq.~(\ref{Eq:Hyperentangled2}) was generated and then measuring the degree of 
entanglement of the output state.  Given a general hyperentangled state with 
arbitrary degrees of entanglement for the polarization and momentum qubit pairs 
of Eq.~(\ref{Eq:Hyperentangled1}), $\theta_P = \theta_M = \theta$ is required 
for the input.  In our verification of the input state, we characterized the 
partially hyperentangled state by comparing the different amounts of 
entanglement in the polarization and momentum modes at different $\theta = 
44^\circ, 41.9^\circ ,39.3^\circ, 35.9^\circ$.  For each $\theta$ the SP 
protocol was carried out on the input state and we analyzed the extracted output 
state.  In particular, we measured how close the extracted state was to the 
expected polarization triplet $|\phi^+\rangle$.

\subsection{Input state characterization}
\label{InputStateCharacterizationSection}
\begin{figure}
\begin{center}
\begin{minipage}{5cm}
	\begin{center}
		\includegraphics[width=5cm]{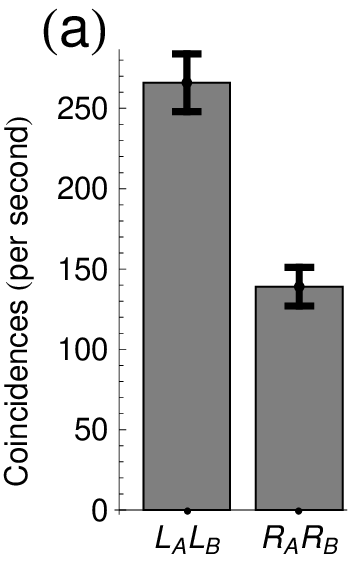}
  \end{center}
\end{minipage}
\
\begin{minipage}{5cm}
	\begin{center}
     \includegraphics[width=5cm]{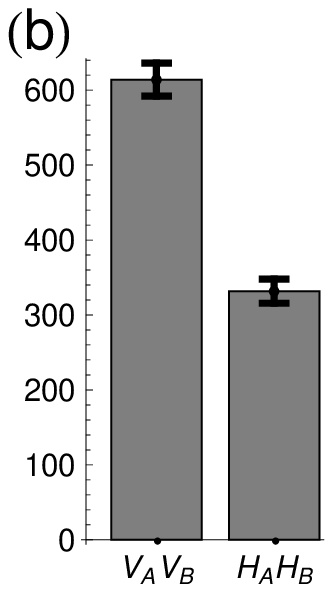}  
  \end{center}
\end{minipage}
\
\begin{minipage}{8cm}
	\begin{center}
     \includegraphics[width=8cm]{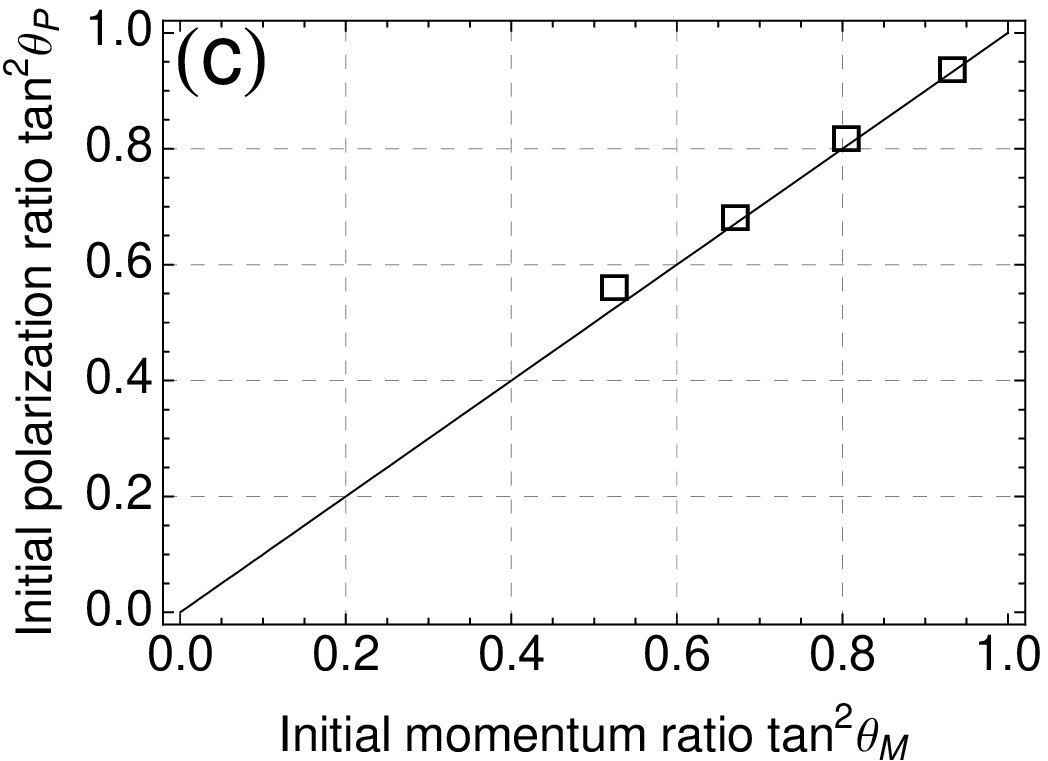}  
  \end{center}
\end{minipage}
\end{center}
\caption{Characterization of initial hyperentangled state. Detected coincidences 
of (a) $|L\rangle_A |L\rangle_B$ and $|R\rangle_A |R\rangle_B$ terms for 
momentum input state $\theta_M=35.9^\circ$ ($\theta_P=45^\circ$), and (b) 
$|V\rangle_A |V\rangle_B$ and $|H\rangle_A |H\rangle_B$ terms for polarization 
input state $\theta_P=35.9^\circ$ ($\theta_M=45^\circ$). (c) Initial state of 
polarization qubits $\tan^2 \theta_P$ as a function of initial state of momentum 
qubits $\tan^2 \theta_M$. Open squares are the measurement results, and the 
straight line indicates the ideal case. Pump power was 5 mW, and detection 
bandwidth was 1 nm. }
\label{DistllationInputPlot}
\end{figure}

To characterize the input hyperentangled state of 
Eq.~(\ref{Eq:Hyperentangled1}), we utilized the SP and {\sc p-cnot} apparatus of 
Fig.~\ref{Fig:DistillSPSetup}(a) and our ability to control $\theta_M$ and 
$\theta_P$ independently, as described in Sect.~III.  The combination of the SP 
and {\sc p-cnot} gate transforms the input state of 
Eq.~(\ref{Eq:Hyperentangled1}) into
\begin{equation}
(\cos\theta_P \sin\theta_M |H\rangle_A |H\rangle_B + 
 \sin\theta_P \cos\theta_M |V\rangle_A |V\rangle_B)
|D\rangle_A |D\rangle_B\,.
\label{Eq:SPtransform}
\end{equation}
To evaluate the distribution of the momentum entanglement, we set 
$\theta_P=45^\circ$ in Eq.~(\ref{Eq:SPtransform}) so that $\cos\theta_P = 
\sin\theta_P = 1/\sqrt{2}$.  As discussed in Sect.~III, we adjusted the relative 
$H$ and $V$ components of the pump to control $\theta_P$.   We then obtained the 
ratio of $\sin^2\theta_M / \cos^2\theta_M$ by measuring the coincidence ratio of 
$C(H, H)$ and $C(V, V)$, where $C(X, Y)$ denotes the detected coincidence rate 
between $X$-qubit measurement by Alice and $Y$-qubit measurement by Bob. 
Any desired momentum entanglement distribution can be obtained by adjusting the SPDC 
crystal temperature while monitoring this coincidence ratio.  
Fig.~\ref{DistllationInputPlot}(a) shows the measured distribution of the 
momentum qubits for $\theta_M = 35.9^\circ$.  
For all the coincidence measurements in this work, each consisted of an average of
60 1-s measurements without background subtraction.

Similarly, to obtain the distribution of the polarization entanglement, we set 
$\theta_M = 45^\circ$ by choosing an appropriate PPKTP temperature. After 
Schmidt projection and the {\sc p-cnot} gate, we measured the polarization 
distribution by monitoring the coincidence ratio of $C(V, V)$ and $C(H, H)$ to 
yield $\tan^2 \theta_P$.  Fig.~\ref{DistllationInputPlot}(b) shows the 
distribution of the polarization qubits for the case of $\theta_P = 35.9^\circ$.  
Note that Fig.~\ref{DistllationInputPlot}(a) and (b) have different scales 
because the two measurements were taken at two different temperatures that 
resulted in different output fluxes.  More importantly is that 
Fig.~\ref{DistllationInputPlot}(a) and (b) show that both ratios are nearly 
identical, as required by the Schmidt projection protocol.  

The above measurements for determining the crystal temperature to obtain 
$\theta_M = 35.9^\circ$ and the pump's $H$--$V$ component ratio to set  
$\theta_P = 35.9^\circ$ were repeated for other $\theta_M$ and $\theta_P$ values 
of $44^\circ, 41.9^\circ$, and $39.3^\circ$.  
Fig.~\ref{DistllationInputPlot}(c) shows the measured polarization coincidence 
ratio $\tan^2 \theta_P$ at various measured momentum coincidence ratio $\tan^2 
\theta_M$, showing clearly that the degree of partial entanglement in the 
polarization and momentum qubit pairs were well matched, 
$\theta_M\approx\theta_P$. The ideal distribution ratio is also shown in 
Fig.~\ref{DistllationInputPlot}(c) as a straight line.  To obtain the partially 
hyperentangled state required for the SP protocol (and for the measurements in 
Fig.~\ref{DistllationInputPlot}(c)), we set $\theta_M = \theta_P = \theta$ in 
the following way.  First, we set 
$\theta_P = 45^\circ$ that could be confirmed by monitoring the coincidence 
ratio in the polarization qubits of the collinear $R$-path outputs. Then we 
adjusted the crystal temperature to the value that would yield the desired 
$\theta_M$, which we verified with the measurements presented in 
Fig.~\ref{DistllationInputPlot}(a).  Finally, we set $\theta_P$ to the same 
value as $\theta_M = \theta$ that we confirmed by again measuring the 
polarization coincidence ratio of the collinear $R$-path outputs.

\subsection{Characterization of Schmidt projected output state}
\begin{figure}
\begin{center}
\begin{minipage}{6cm}
	\begin{center}
		\includegraphics[width=6cm]{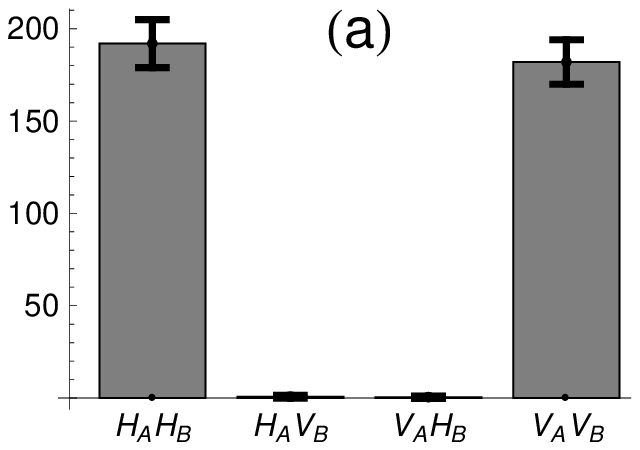}
  \end{center}
\end{minipage}
\
\begin{minipage}{6cm}
	\begin{center}
     \includegraphics[width=6cm]{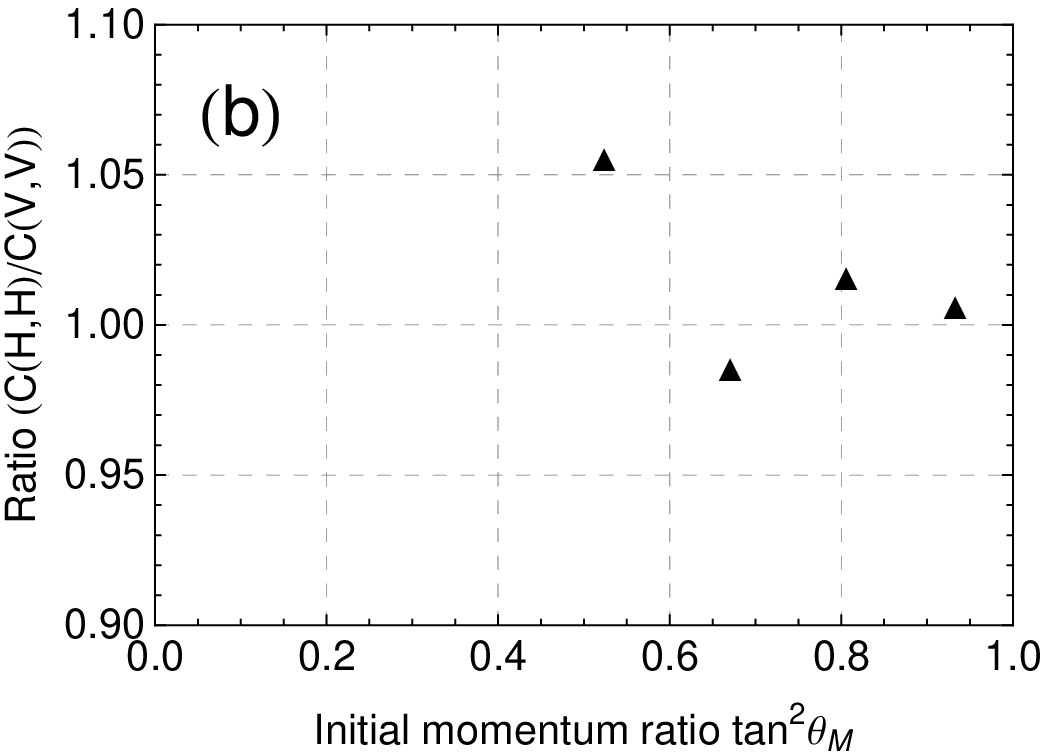}  
  \end{center}
\end{minipage}
\end{center}
\caption{Coincidence measurements of output state in the $H$--$V$ basis. (a) 
Measured coincidence rate of the two parallel and two orthogonal polarization 
terms of the output state for input state with $\theta=35.9^\circ$.  (b) Plot of 
$C(H, H)/C(V, V)$ of the output state as a function of initial momentum qubit 
ratio $\tan^2 \theta_M$. Pump power was 5 mW, and detection optical bandwidth 
was 1 nm.}
\label{DistllationOutputPlot}
\end{figure}

After the distillation procedure, we expect the final state to be a maximally 
polarization-entangled triplet state $|\phi^+ \rangle$ of 
Eq.~(\ref{Eq:triplet}).  We first analyzed the output state by measuring the 
coincidences between Alice and Bob in the $H$--$V$ basis, that for 
$|\phi^+\rangle$ should yield equal probability of detecting parallel 
polarizations but zero probability of detecting orthogonal polarizations.   
Fig.~\ref{DistllationOutputPlot}(a) shows the four measured coincidence rates 
$H_A H_B$, $H_A V_B$, $V_A H_A$, and $V_A V_B$ for an input state with 
$\theta\simeq 35.9^\circ$.
The measurement results clearly show that the output 
state had nearly balanced $|H\rangle_A |H\rangle_B$ and $|V\rangle_A 
|V\rangle_B$ terms and negligible $|H\rangle_A |V\rangle_B$ and $|V\rangle_A 
|H\rangle_B$ components.  We then made the same coincidence measurements for 
input states with different values of $\theta$.  The coincidence rates for 
orthogonal polarizations $H_A V_B$ and $V_A H_B$ were negligible, as expected 
for $|\phi^+ \rangle$.   Fig.~\ref{DistllationOutputPlot}(b) plots the ratio 
of coincidence rates $C(H, H)/C(V, V)$ of the output state (solid triangles) as 
a function of the input state set by $\theta$.  The measured coincidence ratios 
lie close to the expected value of unity for a maximally entangled output state 
with less than 5\% error.

\begin{figure}[h]
\begin{center}
\begin{minipage}{6cm}
	\begin{center}
		\includegraphics[width=6cm]{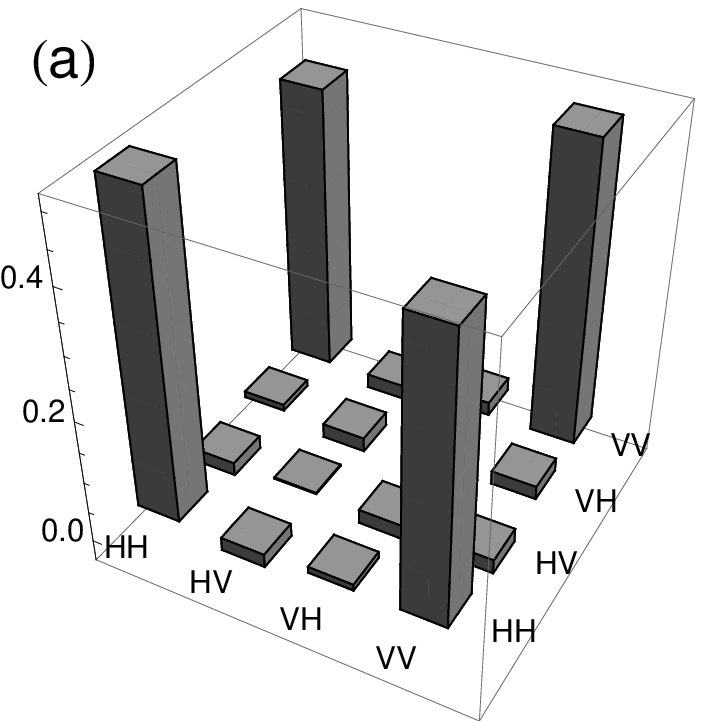}
  \end{center}
\end{minipage}
\
\begin{minipage}{6cm}
	\begin{center}
     \includegraphics[width=6cm]{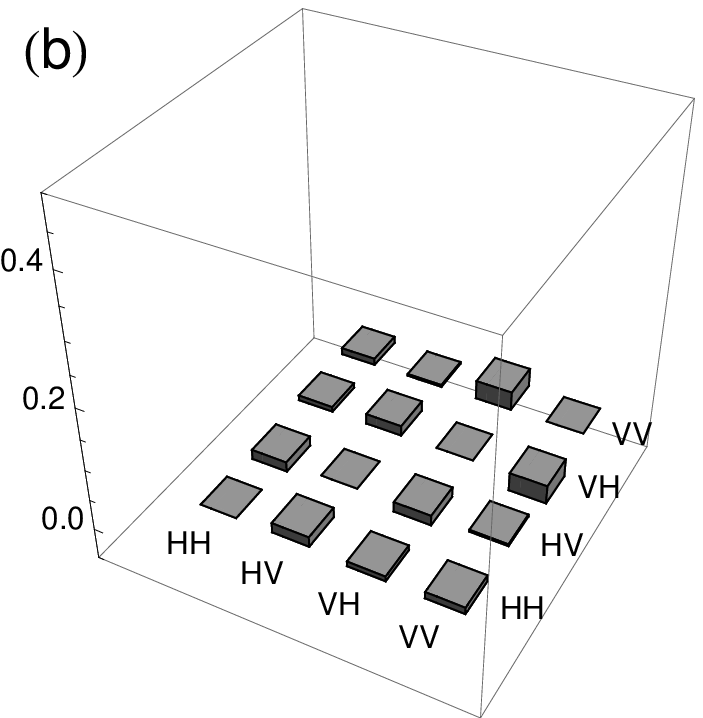}  
  \end{center}
\end{minipage}
\end{center}
\caption{(a) Real part, and (b) imaginary part of measured density matrix of the 
Schmidt projected output state for initial state with $\theta=35.9^\circ$. }
\label{Fig:DistillDensityMatrix}
\end{figure}

To show that the distilled output was indeed entangled as a coherent 
superposition, and not as a classical mixture, of $|H\rangle_A 
|H\rangle_B$ and $|V\rangle_A |V\rangle_B$, we made additional measurements for 
verification.
We first performed quantum state tomography \cite{QTomography} on the output state.
We can reconstruct the density matrix of Alice's and Bob's polarization qubits by
making coincidence measurements for the 16 combinations of their polarization analyzers
set along $|H\rangle$, $|V\rangle$, $(|H\rangle+|V\rangle)/\sqrt{2}$, and $(|H\rangle-i|V\rangle)/\sqrt{2}$.
The real and imaginary parts of the output density 
matrix $\rho$ for the case $\theta = 35.9^\circ$ are displayed in 
Fig.~\ref{Fig:DistillDensityMatrix}(a) and (b), showing clearly that the state 
is $|\phi^+\rangle$ with a fidelity, $\langle\phi^+|\rho|\phi^+\rangle$, of 
0.952.  

\begin{figure}
  \begin{center}
    \includegraphics[width=9cm]{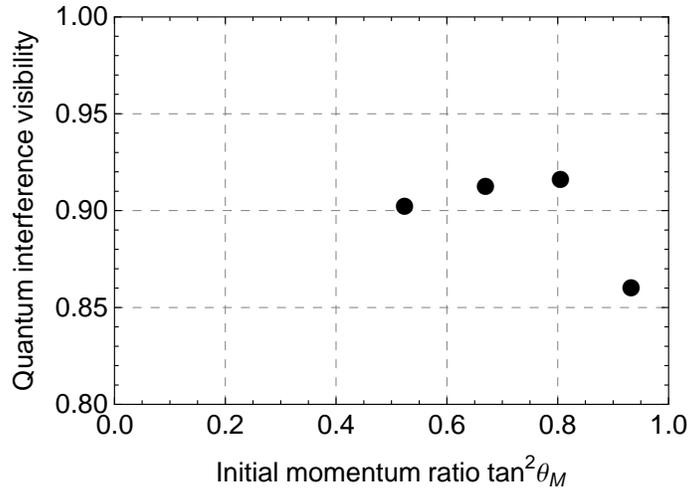}
  \end{center}
  \caption{Measured quantum-interference visibilities in the $\pm 45^\circ$ diagonal basis $(|H\rangle\pm|V\rangle)$.}
  \label{Fig:DistillVisibility}
\end{figure}

Another useful indicator of the output state coherence is two-photon quantum 
interference visibility measured in the $\pm 45^\circ$ diagonal basis 
$(|H\rangle\pm|V\rangle)$. Quantum interference visibility is defined as
$(C_{max}-C_{min})/(C_{max}+C_{min})$, where $C_{max}$ is the maximum
coincidence counts and $C_{min}$ is the minimum coincidence counts.
Fig.~\ref{Fig:DistillVisibility} plots the 
measured visibilities for different input states, all showing $\sim$90\% 
visibility except one lower-visibility point that was caused by a slight 
misalignment of the apparatus.  The visibilities in 
Fig.~\ref{Fig:DistillVisibility} were primarily limited by the imperfect {\sc 
p-cnot} gate that had a classical visibility of only $\sim$93\% \cite{PCNOT}.  
Another possible source of the visibility loss is the slight spectral mismatch 
of the output photons from the $L$ and $R$ output spatial modes.  Even though 
the spectra of the output photons were mainly determined by the two interference 
filters, a slight mismatch in the filtered output spectra could happen because  
slightly different emission angles for the $L$ and $R$ paths (see 
Fig.~\ref{Fig:DistillGeneration}) would yield slightly different spectral 
distribution.

\begin{figure}
  \begin{center}
    \includegraphics[width=9cm]{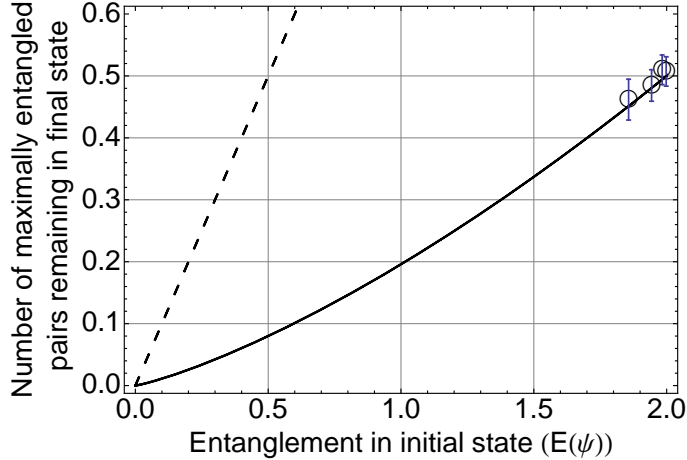}
  \end{center}
  \caption{Measured efficiency of Schmidt projection implementation.  
Expected number of maximally entangled pairs after Schmidt projection is 
applied to a pair of hyperentangled state as a function of 
entanglement in the initial 
state ($E(\psi)$).  Open circle is the measured number and the solid curve 
is the theoretical prediction when Schmidt projection is applied to a 
hyperentangled state.  The dashed curve shows the maximum number of maximally entangle pairs
obtainable by Schmidt projection when the input consists of an  infinite number 
of pairs.}
  \label{Fig:DistillEffPlot}
\end{figure}

From the different state analysis measurement results shown in 
Fig.~\ref{DistllationOutputPlot}, \ref{Fig:DistillDensityMatrix}, and 
\ref{Fig:DistillVisibility}, we conclude that our distillation outputs were in 
the $|\phi^+\rangle$ state independent of the input states with different 
$\theta$ values.  One of the key advantages of the SP protocol is its high 
distillation efficiency, especially for a large number of initial pairs.  We 
have measured the efficiency of our Schmidt projection implementation in terms 
of the expected number of maximally entangled pairs remaining in the final state
as we varied the amount of initial entanglement $E(\psi)$ that we plot
in Fig.~\ref{Fig:DistillEffPlot} as open circles. We find that the measured efficiency
is in good agreement with the theoretically calculated values (solid curve).
For comparison, the dashed line in Fig.~\ref{Fig:DistillEffPlot} shows the maximum number of maximally 
entangled pairs obtainable by the Schmidt projection protocol when 
it is applied to an infinite number of input pairs.  In general, Schmidt 
projection shows a lower yield than Procrustean methods \cite{Concentration, 
Procrustean} when it is applied to less than 5 pairs, as in our case. However, 
even with a small number of initial pairs, Schmidt projection still has the 
practical advantage that it is not necessary to adjust the distillation setup as 
a function of the input states.

\section{DISCUSSION}
In this work, we have experimentally demonstrated the Schmidt projection method
that is considered one of the most powerful distillation protocols.  We 
employed hyperentangled photons produced by spontaneous parametric 
down-conversion to generate pairs of partially-entangled states.  We developed a 
technique to independently control the amount of entanglement in the 
polarization (by pump adjustments) and momentum (by crystal temperature) degrees 
of freedom of the hyperentangled photon pairs.  This entanglement 
control capability was utilized to prepare the identical pairs of partially entangled qubits 
as the initial state for the SP protocol.  Schmidt projection on the 
hyperentangled photon pairs was realized by a PBS and two {\sc m-cnot} gates, 
and we used a {\sc p-cnot} gate to convert the output into a polarization Bell 
state.  Our experimental results show that Schmidt projection can distill a pair 
of maximally entangled qubits in the form of a polarization-entangled triplet 
and it is independent of the degree of entanglement in the initial state.  
Furthermore, the measured distillation efficiencies agree with the theoretical 
prediction.  We note that Schmidt projection can be selectively applied to 
perform entanglement purification.  If classical communication is used to 
compare the projection results, then certain types of mixed states can be 
purified, such as $\rho=\Sigma p|\psi\rangle\langle\psi| + (1-
p)|01\rangle\langle01|$, as well as a mixture of two identically decohered pairs 
\cite{Parity1}.  

\begin{figure}
  \begin{center}
    \includegraphics[width=9cm]{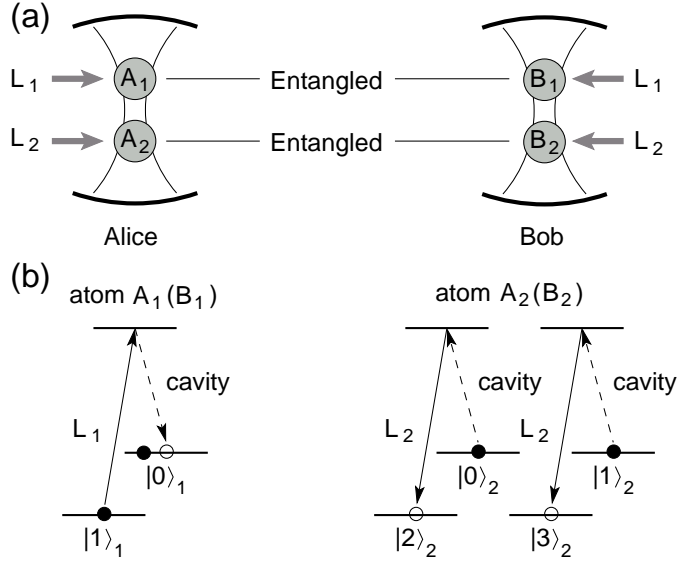}
  \end{center}
  \caption{Schematic illustration of applying Schmidt projection to cQED\@.  (a) 
Alice and Bob have two pairs of entangled atoms (A$_1$--B$_1$, A$_2$--B$_2$) 
trapped in an optical cavity. Pulses from two lasers, L$_1$ and L$_2$, are 
applied in a specific sequence to implement adiabatic transfer of the dark state  
\cite{cQED}. (b) Proposed electronic-level structure for implementing the SP 
protocol. Atoms A$_1$ and A$_2$ are of the same kind (e.g., Rb 
\cite{BellMeasurement}), and different levels are used to encode the initial 
qubit state.  After adiabatic passage, A$_1$ will be in the $|0\rangle_1$ state, 
and A$_2$ will be a superposition of all four states 
($|0\rangle_2$--$|3\rangle_2$).}
  \label{Cavity}
\end{figure}

It is important to recognize that the key to our Schmidt projection 
implementation is that distinct tensor products of local qubits, 
$|0\rangle_1|0\rangle_2$, $|0\rangle_1|1\rangle_2$, $|1\rangle_1|0\rangle_2$, 
and $|1\rangle_1|1\rangle_2$, are encoded in physically distinguishable states, 
$|VL\rangle$, $|VR\rangle$, $|HL\rangle$, $|HR\rangle$.  Therefore the same 
concept can be applied to other qubit systems such as cavity quantum 
electrodynamics (cQED) \cite{cQED,BellMeasurement} or trapped ion 
\cite{IonTrapCNOT} systems, where multiple-qubit product states can be mapped 
into different internal states within the same atom.  For example, in a cQED 
system illustrated in Fig.~\ref{Cavity}(a), Alice has two trapped atoms each of 
which is entangled with the corresponding trapped atom in Bob's cavity, where 
the qubits are stored in the ground-state Zeeman levels.  Consider the case in 
which the first qubit is stored in $|0\rangle_1$ and $|1\rangle_1$ of A$_1$ 
(B$_1$) and the second qubit is in $|0\rangle_2$ and $|1\rangle_2$ of A$_2$ 
(B$_2$) on Alice's (Bob's) side as shown in Fig.~\ref{Cavity}(a).  One can 
coherently transfer the quantum states stored in Alice's two atoms completely to  
A$_2$ while leaving A$_1$ in $|0\rangle_1$ by adiabatic passage via a dark state 
of the two-atom + cavity system \cite{cQED}.  That is, $|p\rangle_1|q\rangle_2 
\Rightarrow |0\rangle_1|p\,q\rangle_2$ when $p\,q$ is in a binary 
representation. By mapping all four possible quantum states into four internal 
states of A$_2$, the undesirable states such as $|0\rangle_2$ or $|3\rangle_2$ 
can be eliminated by driving the cycling transitions from the undesirable states 
to an auxiliary state (not shown in the level structure) \cite{BellMeasurement}, 
and the process can be confirmed by the induced fluorescence. If no fluorescence 
is observed, we can conclude that Schmidt projection has successfully taken 
place. 

In the trapped-ion case, one can use the quantized center-of-mass vibrational 
mode (`bus mode') in a similar way as in the cQED case.  
When there are two ions, the electronic state of the first ion can be 
transferred to the bus mode \cite{IonTrapCNOT} with a red-sideband transition, 
and this bus mode can be transferred to the second ion while mapping four 
different two-qubit states to four distinct internal states of the second ion. 
The rest of the procedure is similar to the cQED case. We believe that further 
studies of Schmidt-projection implementations in atoms and ions and for mixed 
states should enhance our expanding collection of tools in quantum information 
science.

This work was partially supported by the Hewlett Packard-MIT Alliance.

\end{document}